# Endurance – Write Speed Tradeoffs in Nonvolatile Memories

Dmitri B. Strukov

Electrical and Computer Engineering Department, UC Santa Barbara

Email: strukov@ece.ucsb.edu

**Abstract**

We derive phenomenological model for endurance-write time switching tradeoff for nonvolatile memories with thermally activated switching mechanisms. The model predicts linear to cubic dependence of endurance on write time for metal oxide memristors and flash memories, which is partially supported by experimental data for the breakdown of metal-oxide thin films.

**Introduction**

Nonvolatile memory devices require very strong nonlinearity in switching kinetics with respect to applied write stimulus (e.g., applied voltage), which is necessary to combine long retention with fast intrinsic write speed [Yan13]. For example, the retention-to-write-speed ratio is more than 10 orders of magnitude for nonvolatile NOR/NAND flash memories (e.g. with 10 years of retention and microsecond/millisecond write time), while it is less than 6 orders of magnitude (tens of milliseconds retention and tens of nanoseconds write time) for volatile dynamic random access memories (DRAM) [ITRS]. Given typical strong nonlinearity in switching mechanism, a very natural way to enhance this ratio in nonvolatile memories is to increase write speed by applying larger stress; however, larger stress is more harmful for device endurance, because the kinetics of the failure mechanisms is also typically strongly enhanced by larger write stress.

For example, in flash devices, the nonlinearity is due to (super) exponential dependence of tunneling current on the applied electric field across gate/floating gate oxides. To get high retention-to-write-speed ratio, the electric fields are typically close to 10 MV/cm [Lik98], which is already very close to typical dielectric breakdown values in $SiO_2$ [McP03]. Increasing write (program/erase) voltages, in principle, will super-exponentially decrease write time [Lik98], however, it will also increase exponentially probability of creation of and/or filling of existing deep traps [McP03], which are the primary reason for limited endurance in flash memories.

A similar endurance-write-speed tradeoff exists for the emerging nonvolatile memories. In valence-change, electrochemical, and thermochemical memories [Was09] (which are often called ReRAM [ITRS] or memristors [Chu11]) the write speed is very often super-exponential function of applied voltage [Yan13] – see for example, experimental evidence of that for valence change memories in Refs. [Ali12, Pic09]. The primary memory mechanism is due to ion profile modulation, such as motion of oxygen vacancies in valence-change and thermochemical memories, or motion of electrode material ions in electrochemical memories. The ionic motion in solids is thermally activated process with exponential dependence on the electric field [Mot40], and, therefore, whenever Joule heating is involved, the switching kinetics is strongly super-exponential [Yan13]. Though there are limited studies of the failure mechanisms in memristive devices, at least preliminary evidence shows that one of the mechanisms is related to the overstressing devices upon write operation. Very high temperatures combined with high electric field can cause melting of the electrode material (or a formed filament) inside the insulating matrix causing permanent failure. Given the thermally activate nature of the failure mechanisms, it is natural to expect that endurance would decrease (super-) exponentially with applied stress.

Endurance failure mechanisms have been studied extensively in a past for flash memories [Cap99] and recently in metal oxide resistive switching devices [Lee10, Che11]. However, to the best of our knowledge, the endurance-write-speed tradeoff were never studied analytically and the main goal of this paper is to derive such model. Our particular goal is to study catastrophic failures, such as e.g. electrode stability or oxide degradation [Che11], as oppose to the endurance problems which can be recovered from by reconditioning the device (e.g. resulted from repetitively applying asymmetric set/reset stress when switching the device on and off).

## Phenomenological Model

To derive endurance-write-speed tradeoff let us assume a model loosely based on the one presented in Ref. [Str09]. In particular, let us assume that in case of memristors (flash memories) to switch memory state of the device an ion (electron), must travel the distance $d$ across insulating matrix (gate oxide) upon application of write voltage $V$. Assuming thermally activated motion with activation energy $U_S$ and local thermal temperature $T$, and constant electric field, the switching speed can be approximated as

$$t_S \approx \frac{d}{v_S} \approx \frac{2d}{fa} \exp[\frac{U_S}{kT}]\exp[-\frac{Vq}{2kT}\frac{a}{d}] , \qquad (1)$$

where $v_S$ is a speed of ions (electrons), $k$ is a Boltzmann constant, $q$ is an elementary charge, $f$ is the frequency of escape attempts, and $a$ is an average hopping distance [Str09]. The temperature is also a function of voltage when Joule heating is involved, i.e.

$$T\text{- 300K} \propto IV \qquad (2)$$

Let us assume that the failure mechanism is due to thermally activated motion of ions (electrons) across the distance $d$ but with higher activation energy $U_F$, so that average time to failure is

$$t_F \approx \frac{d}{v_F} \approx \frac{2d}{fa} \exp[\frac{U_F}{kT}]\exp[-\frac{Vq}{2kT}\frac{a}{d}] . \qquad (3)$$

With such definition, the endurance, i.e. the number of times the devices can be switched before the failure occurs, is proportional to the ratio of failure to switch times:

$$\text{endurance} \approx t_F/t_S . \qquad (4)$$

Figure 1a shows endurance-write-speed tradeoff data calculated with Eqs. 1- 4 for several plausible temperature-voltage dependences (Fig. 1b). The considered temperature dependences are phenomenological and based on typically observed local temperatures and possible heating mechanisms. For example, a quadratic dependence is representative of ohmic conductance, while exponential dependence is representative of exponential tunneling current transport [Bor09]. For all cases, it is assumed that the local temperature is 300K at 0 V and 600K for 1V applied voltage.

In particular, Figure 1a shows that if the activation energy for failure mechanism is not too high and only by 1 eV larger than that of switching, which would be representative of low endurance devices, the increase in speed is linearly proportional to the decrease in endurance. For larger $U_F$ the increase in speed results in approximately quadratic and cubic drop in endurance for $U_F$ = 3 eV and 4 eV, correspondingly. The results show that the tradeoff is mostly determined to the difference between $U_F$ and $U_S$ and almost insensitive to the particular law of temperature dependence and other parameters. Moreover, for relatively small electric field $Vqa << 2dU_S$ the following approximate formula can be used:

$$\text{endurance} \approx \left(\frac{t_S}{t_0}\right)^{\frac{U_F}{U_S}-1},\qquad(5)$$

where $t_0 = 2d/(fa)$.

**Discussion and Summary**

Assuming reasonable values of $d$ = 10 nm, $a$ = 0.2 nm, and $f = 10^{13}$ Hz results in $t_0 = 10^{-11}$ s. Using the same values, Figure 2 shows how write delay scales with temperature and voltage for two values of activation energies. Note that though temperature is a function of voltage, it is still convenient to consider it as independent parameter given broad range of temperature-voltage relations. From Figure 2a, $T_S$ = 100 ns implies ~650 K internal temperature at ~1 V applied voltage. With $U_F$ = 2 eV the endurance is of the order of 10,000 for this case. If write speed is increased by factor of 100, e.g. by applying 2 V and causing internal heating up to ~720 K, the endurance is decreased accordingly to 100. For more robust devices, e.g. assuming $U_F$ = 3 eV and the same other parameters, the corresponding values of endurances are $10^8$ and $10^4$, respectively.

Because of similar physics the model can be applicable to flash memory devices. For example, results in Ref. [Xiu07] support the proposed tradeoff for flash memories with parameters in Eq. 5 close to the ones considered for memristive devices in the discussion above. In particular, Figure 3 shows that exponent in Eq. 5 ranges from 1 to 2, while $t_0$ parameter ranges from $10^{-13}$ to $10^{-15}$ for different types of oxides. The tradeoff might be also similar for many other emerging nonvolatile memories based on phase-change [Rao12], magnetoresistive, and ferroelectric materials [Tsy12], because of thermally activated nature of switching. For example, write kinetics was studied in phase change memories [Rao12] and it was shown that switching speed depends exponentially on the applied electric field and temperature [Kar08].

It is worth noting that in memristors with filamentary switching, there exists at least one other possible mechanism which can lead to higher endurances with larger stresses, which is the opposite of the discussed tradeoff. In such memories, the smaller stress can in principle cause more unwanted lateral diffusion away from the filament relative to the useful (i.e. in terms of switching) drift along the filament. Understanding the importance of this effect as compared to the considered failure mechanisms will require more experimental and theoretical studies.

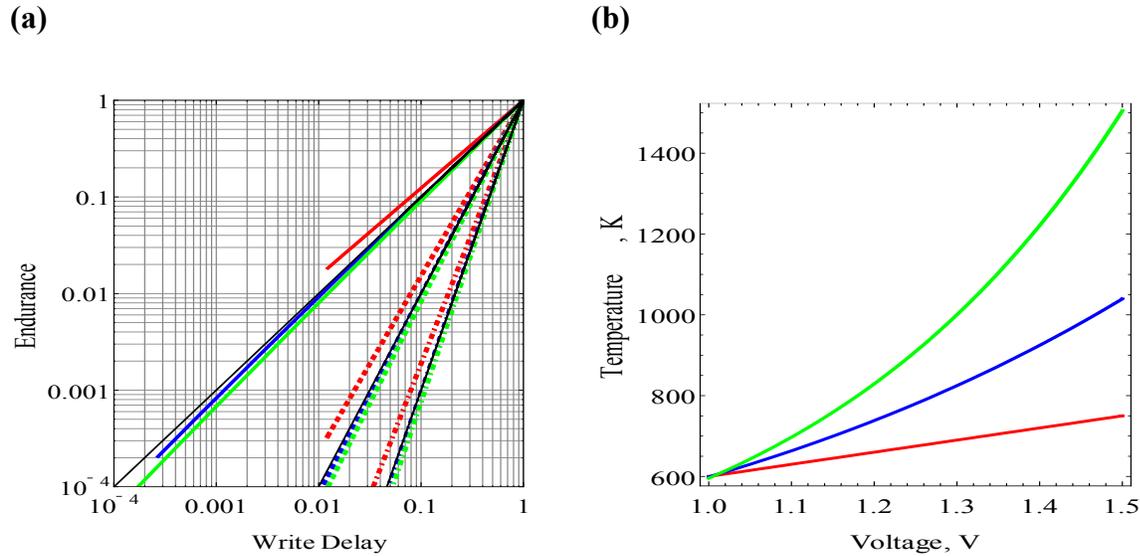

Figure 1. (a) Normalized endurance vs. normalized write delay for three values of $U_F$ = 2 eV (solid lines), 3 eV (dashed), and 4 eV (dot-dashed) and (b) three functions of $T = f(V)$ – i.e. linear (red), exponential (blue) and steep exponential (green). Other assumptions are that $U_S$ = 1 eV, $a/d$ = 0.1 and that applied voltage $V$ is changing from 1 V to 1.5V. Both write delay and endurance for every particular tradeoff curve are normalized with respect to the slowest write time and the highest endurance. Solid black lines are just guides for an eye showing linear, quadratic and cubic endurance-write-delay relations.

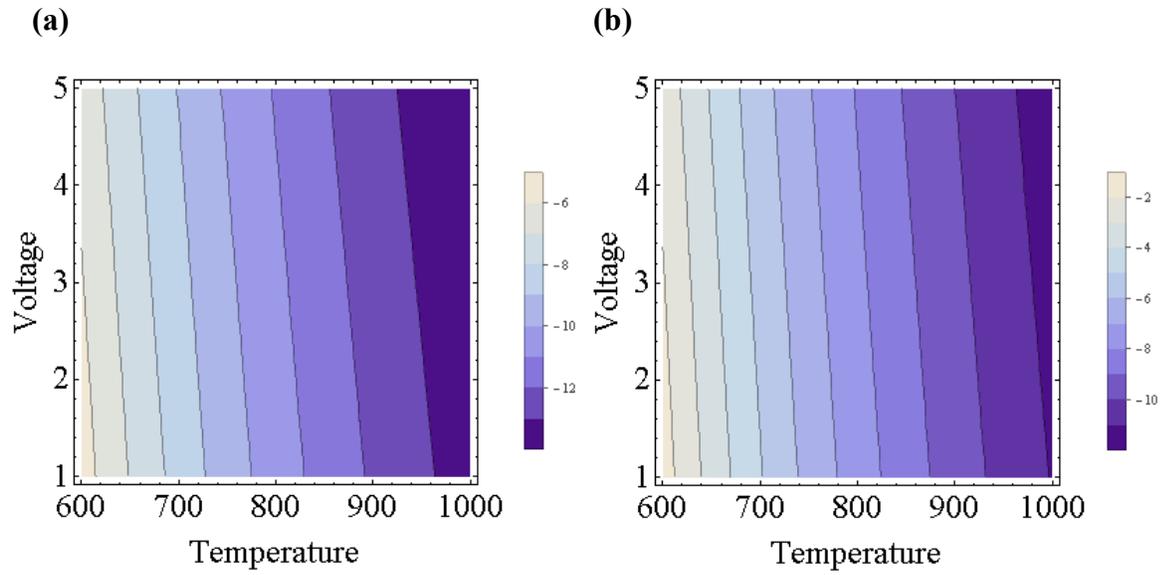

Figure 2. Estimated write delay (shown in seconds on a log scale) as a function of internal temperature and applied voltage for (a) $U_S$ = 1.0 eV and (b) $U_S$ = 1.2 eV and $d$ = 10 nm, $a$ = 0.2 nm, and $f$ = $10^{13}$ Hz.

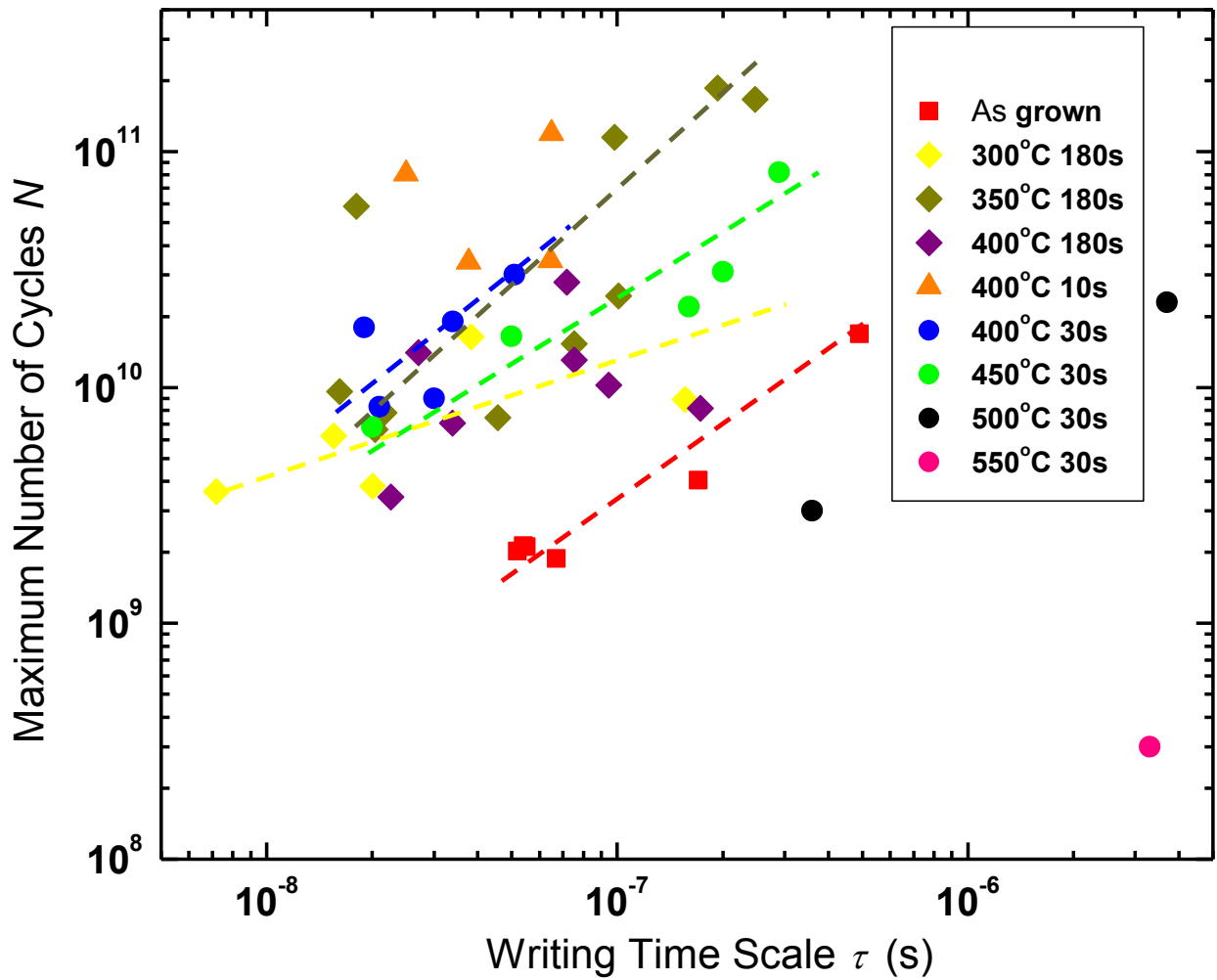

Figure 3. Experimental data for rf-plasma grown AlO$_x$ barrier endurance as a function a voltage applied to the barrier, which is recalculated as a corresponding write time on x axis. Reproduced from Ref. [Liu07]. The dashed lines are power law fitting using Eq. 5.